\documentclass[a4paper]{PoS}

\title{Prospects for Measuring the Positron Excess with the Cherenkov Telescope
Array}

\ShortTitle{Measuring the Positron Excess with CTA}

\author{\speaker{Justin Vandenbroucke}\\
        University of Wisconsin, Madison\\
        E-mail: \email{vandenbrouck@wisc.edu}}

\author{Peter Karn\\
        University of Wisconsin, Madison\\
        E-mail: \email{peter.j.karn@gmail.com}}

\author{Matthew Wood\\
        SLAC National Accelerator Laboratory\\
        E-mail: \email{mdwood@slac.stanford.edu}}

\author{Pierre Colin\\
        Max-Planck-Institut f\"ur Physik, Munich\\
        E-mail: \email{colin@mppmu.mpg.de}}

\author{for the CTA consortium\footnote{Full consortium author list at
http://cta-observatory.org}}

\abstract{The excess of positrons in cosmic rays above $\sim$10 GeV has been a
  puzzle since it was discovered.  Possible interpretations of the excess have
  been suggested, including acceleration in a local supernova remnant or
  annihilation of dark matter particles.  To discriminate between these
  scenarios, the positron fraction must be measured at higher energies.  One
  technique to perform this measurement is using the Earth-Moon spectrometer:
  observing the deflection of positron and electron moon shadows by the Earth's
  magnetic field.  The measurement has been attempted by previous imaging
  atmospheric Cherenkov telescopes without success.  The Cherenkov Telescope
  Array (CTA) will have unprecedented sensitivity and background rejection that
  could make this measurement successful for the first time.  In addition, the
  possibility of using silicon photomultipliers in some of the CTA telescopes
  could greatly increase the feasibility of making observations near the moon.
Estimates of the capabilities of CTA to measure the positron fraction using
simulated observations of the moon shadow will be presented.}

\FullConference{The 34th International Cosmic Ray Conference,\\
		30 July- 6 August, 2015\\
		The Hague, The Netherlands}

\begin{document}

\section{Introduction}

Positrons can be created by interactions between cosmic rays and the
insterstellar medium.  Our knowledge of the cosmic ray spectrum and the
interstellar environment allows us to predict the spectrum of positrons incident
upon the Earth due to this secondary production of positrons.  Since the 1970s
there have been experimental hints in the measured positron fraction and the
total electron and positron flux that the spectrum of positrons at Earth
deviates from that expected from secondary production, indicating additional
production of positrons \cite{hints}.  The first definitive measurement showing
that the positron fraction is rising at energies >1 GeV came from the PAMELA
experiment in 2009 \cite{pamela}.  More recently, this result has been confirmed
by the Fermi-LAT \cite{fermi} and extended up to 500 GeV by the AMS-02
experiment \cite{ams}. 

There are several interpretations of the rising positron fraction at high
energies which correspond to different sources of the positrons
\cite{interpret}.  One example is the production of secondary positrons inside
supernova remnants \cite{model}.  Such positrons would have a harder spectrum
than those produced through interactions with the interstellar medium, and
therefore could produce the anomaly in the positron fraction.  Another
possibility is the presence of a nearby object that is accelerating positrons
such as a pulsar or microquasar.  A third possibility is the production of
electrons and positrons due to the annihilation or decay of dark matter
particles.  This is a very exciting idea but there are many models of dark
matter.  There is quite a bit of uncertainty in both the parameter space of dark
matter and the astrophysical models but some distinguishing characteristics have
been predicted.  For example, an important clue is the exact shape of the
positron fraction at high energies.  If the positrons are due to dark matter,
the positron spectrum should have a sharp cutoff at an energy determined by the
dark matter mass, which is then smoothed by propagation effects.  There must be
a high energy limit for secondary positron acceleration in astrophysical sources
as well, but the cutoff should be less sharp.  To address this question, a
measurement of the positron fraction beyond the energy reach of AMS-02 could be
crucial (current publications report the electron flux up to 700 GeV and the
positron flux up to 500 GeV).

The positron fraction can be measured at high energies using what is known as
the Earth-Moon spectrometer.  A fraction of cosmic rays that would have arrived
at Earth are blocked by the moon creating a deficit in the distribution of
cosmic-rays as observed from the Earth (a moon shadow).  The subsequent
deflection of the cosmic rays by Earth's magnetic field as they travel from the
moon to Earth causes the moon shadow to shift in different directions for
positively and negatively charged particles. In a certain energy range, which
depends somewhat on the local magnetic field, the particles are bent enough that
the two shadows are fully separated but not so much that the shadows are smeared
out over a large section of the sky so that the shadows are observable.  By
binning the events in energy, it should be possible to resolve the moon shadows
of particles at different energies and therefore measure the spectrum of
electrons and positrons.  Imaging atmospheric Cherenkov telescopes (IACTs) are
well suited for this measurement because of their good angular and energy
resolution.  The use of this technique to measure the positron fraction was
first attempted with the ARTEMIS experiment, but the background conditions
proved too challenging \cite{artemis}.  More recently, this measurement has been
attempted with the MAGIC observatory which is an ongoing effort \cite{magic}.  

The measurement of the positron fraction using the Earth-Moon spectrometer is
challenging for several reasons.  IACTs are designed to be operated in low
moonlight conditions because photomultiplier tubes (PMTs) are damaged by ambient
light.  This severely limits the ability to point the telescope to a location
near the moon (as close as possible for the highest energy measurements).
Observing during low moon phase is favorable, but observing is also only
possible when the source is relatively high in the sky, which for the moon tends
to be at high phase.  This typically means that a few tens of hours are
available for this observation per year.  The ARTEMIS measurements were made
using special filters designed to attenuate moonlight while allowing the
Cherenkov light through, but the background light conditions were nonetheless
challenging.  The other principal background that complicates this measurement
is protons.  Generally, even after standard gamma-hadron separation cuts have
been applied, the number of residual protons will still outnumber positrons of
the same energy by an order of magnitude.  Since protons and positrons have the
same charge, they will be deflected nearly identically at ultrarelativistic
energies, and therefore the positron moon shadow at a given energy will be
hidden in a proton shadow at that same energy.  Additional work must be done in
data analysis to extract the positron signal from an observation of the positive
shadow.  Observation of the negative shadow, where electrons have a deficit and
protons and positrons are distributed uniformly will give another handle on the
proton contribution.

The Cherenkov Telescope Array (CTA) is a next-generation IACT which is currently
in the planning and prototyping stage and scheduled to finish construction in
2020 \cite{cta}.  CTA will have an array in both the North and South hemispheres
and each array will consist of many tens of telescopes instead of two to five as
in the current generation.  Current plans for the southern array could feature
over 100 telescopes spread over an area 3 km in diameter.  By using three
different telescope sizes, it will be able to cover a wide energy range, from a
few tens of GeV to a few hundreds of TeV with an order of magnitude improvement
in sensitivity over current IACTs.  CTA will have a better chance of
successfully measuring the positron fraction than any previously conceived IACT
for several reasons beyond the overall increase in area and sensitivity.  First,
CTA will have a wide field of view, around $8^{\circ}$ for the medium-sized
telescopes.  This is critical for this study because observing the shadow at
different angular separations away from the moon is equivalent to measuring
different energies.  A wider field of view means the flux can be determined in
more energy bins in one pointing.  More than one pointing may not be possible
because as stated above, there will not be many hours available for the
measurement (especially with the competition of many other exciting observing
targets).  Another reason CTA will be superior to previous IACTs for this
measurement is that some of the camera designs feature silicon photomultipliers
(SiPMs) rather than the traditional PMTs.  All of the small-sized telescopes and
some of the medium-sized telescopes are proposed to be instrumented with SiPMs,
which are much more robust than PMTs and will not be damaged by the high
moonlight conditions.  The large number of telescopes in CTA makes it
impractical to temporarily install filters, but the SiPMs will allow those
telescopes to be pointed at or near the moon.  There will be an increase in
energy threshold for such observations, but no threat to the photodetectors.
This has been proven by the FACT telescope, which has observed showers while
tracking the full moon using a SiPM camera \cite{fact}.  Finally, improvements
in camera technology and a greater telescope multiplicity will allow for
improved gamma-hadron separation, which will make the task of rejecting the
large proton background easier.  The remainder of this paper will describe the
results of a proof-of-concept study to show the capabilities of CTA to measure
the electron and positron spectra using simulations, with a focus on a method
for the rejection of the proton background.

\section{Models, Simulations, and Assumptions}

To evaluate the capabilities of CTA to detect electrons and positrons, a model
for the spectra of electrons and positrons at high energies was needed.  In a
recent paper by Mertsch and Sarkar, they present calculations based on the
supernova remnant hypothesis of secondary positron acceleration \cite{model}.
They use recent data from AMS-02 to fix some of the free parameters of their
model and extrapolate the spectra to high energies (as well as make predictions
about other cosmic ray species).  Within the remaining allowed variations in
model parameters, there is the possibility that the positron fraction could rise
to nearly 50\% at a few TeV, making this model useful for studying the
performance of our analysis method.

Next, it is necessary to simulate the response of CTA to particles with the
above spectra.  The number of each type of telescope has not been finalized, nor
has the exact layout, but simulations of various array configurations have been
carried out and the results published \cite{mc}.  For this project, the array
configuration designated as `2A' from the second round of Monte Carlo production
(i.e. Prod-2) was used, which consists of 4 Large-Sized Telescopes (LSTs), 24
Medium-Sized Telescopes (MSTs), and 37 Small-Sized Telescopes (SSTs).  It is worth
noting that the number of SSTs is likely to be considerably more than what is
simulated here, while not all the MSTs will be instrumented with SiPMs.  In
addition, the level of night sky background in these simulations was fairly low,
corresponding to the brightness of an extragalactic field with an integral
intensity of 0.22/ns/sr/cm$^2$ between 300 and 650 nm.  Nevertheless, the
simulation yields sufficiently reasonable quantities to begin this study.  Using
the resulting simulated effective area and residual protons as a function of
energy, the expected rate of electrons, positrons, and protons in chosen energy
bins can be calculated for any input spectrum.  

To produce a simulated observation of the positive and negative moon shadows, we
needed to model the deflection of the particles by the Earth's magnetic field.
The deflection will depend on the local magnetic field at the location of the
array, which is still not known as of this writing.  In the interest of
comparing to existing studies, it was decided to create our simulations using an
approximation previously used by the MAGIC collaboration for their site on La
Palma, i.e. the deflection is $~1.5^{\circ}(\textrm{TeV}/E$).  This means
the highest energy accessible with this measurement is 3 TeV, where the shadows
would be displaced by $0.5^{\circ}$, or 1 moon diameter.  In this simulation a
coordinate system is chosen such that the true moon location is at the origin
and the magnetic deflection is along the x-axis.  Then for each energy bin, the
shadow covers a range of x-coordinates.  There is also spread in both the x and
y directions by factoring in the point spread function, which is larger for the
residual protons than for the leptons (also obtained from simulation).  

In our simulation, there are 9 energy bins extending from 100 GeV up to about 3
TeV (6 bins per decade).  Each energy bin and shadow (one to the left of the
moon and one to the right) is treated as a separate 100 hour observation with
spatial boundaries corresponding to a rectangle which is 5 times the proton PSF
away from the location of the moon shadows at either end of the energy bin.
These simplifications were chosen to give a clear and complete image of the
shadows to the analysis so it could be tested without worrying about edge
effects, but plans for a more realistic simulation are outlined at the end
of these proceedings.  The expected number of electrons, positrons, and protons
for the 100 hour duration and appropriate solid angle are thrown uniformly over
the area.  It is determined whether or not each one would have been blocked by
the moon based on its deflection, and the PSF smearing is applied.  In this way
a simulated observation of the moon shadows in each bin is generated consisting
of a uniform flux with a deficit of the appropriate shape, location, and depth.
It is assumed that the measurement is done on an area of the sky that is not
near any source of gamma rays.  The diffuse gamma ray background is at least two
orders of magnitude lower than the flux of cosmic rays, so only electrons,
positrons, and protons are simulated.

\section{Template Method}

As stated above, a significant challenge of this measurement is removing the
residual protons.  Most protons can be removed from the data by analysis of the
shape of the shower in the image.  CTA will have a higher efficiency in this
process than any previous IACT.  However, some proton showers are more difficult
to reject due to the stochastic nature of the shower development.  For example,
a proton shower can create an electromagnetic subshower far from the rest of the
shower which would then be indistinguishable from a shower resulting from a
gamma ray, electron, or positron.  Due to the overwhelming prevalence of protons
in cosmic rays, these residual protons will still outnumber positrons by an
order of magnitude in CTA data, according to our simulations.

\begin{figure}
  \centering
  \includegraphics[width=0.8\textwidth]{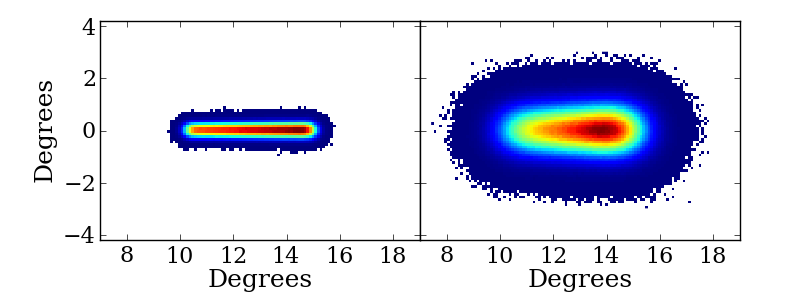}
  \caption{Comparison of templates for the positron deficit on the left and the
    proton deficit on the right for the energy bin from 100 to 147 GeV.  The
    origin is placed at the actual moon location.  The color scales are
    independently normalized.  The difference in PSF for the two particles
  allows the use of the template method to remove the proton background.}
  \label{tempComp1}
\end{figure}

\begin{figure}
  \centering
  \includegraphics[width=0.8\textwidth]{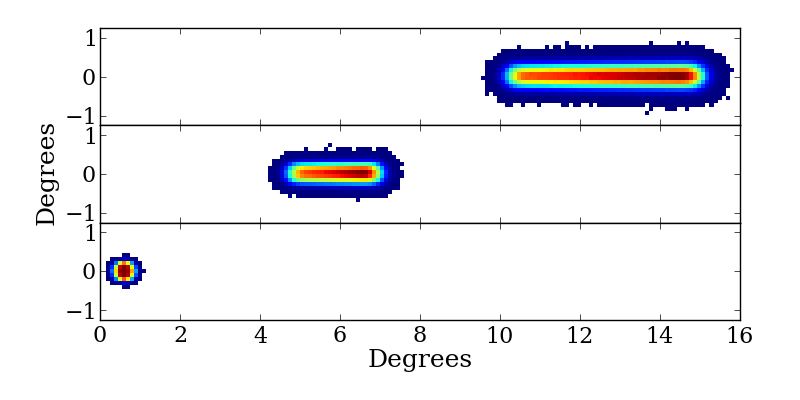}
  \caption{Comparison of templates for the positron deficit for three different
    energy bins: 100-147 GeV on the top, 215-316 GeV in the middle, and 2-3 TeV
    on the bottom.}
  \label{tempComp2}
\end{figure}

At ultrarelativistic energies, the deflection of positrons and protons is nearly
identical so the shadows for both particles lie on top of each other.  However,
the proton shadow is much wider because the PSF for protons is larger.  The
template method is a technique which exploits this fact to remove the protons.
Two-dimensional templates for the shapes of the shadows of different particles
are simulated by assuming the spectra can be approximated as power laws within
an energy bin.  Then particles are thrown from that power law, deflected, and
smeared with the appropriate PSF as above.  The templates are broken up into
small pixels ($0.1^{\circ}$) so that the shapes of the shadows can be accurately
distinguished.  Different power law indices can be simulated and fit as
hypotheses.  The normalizations of the three components are fit to the data
using a $\chi^2$ minimization which gives as output the fluxes of each particle
and associated errors.  The two shadows are fit simultaneously with the
appropriate templates used for each shadow.  For example, for the "positive"
shadow, the electrons are fit using a flat template, the protons are fit using
the wider deficit template, and the positrons are fit using the narrower deficit
template.  The template used for the positron deficit on one side of the moon
and the electron deficit, while similar in shape, must be different because the
two species will generally have different power law indices in a given bin,
which changes the depth of the deficit.  Some examples of templates are shown in
Figs. \ref{tempComp1} and \ref{tempComp2}.

To improve the fit, the total electron-positron spectrum can be used as a prior.
The prior could be based on the spectrum derived from CTA or a previous
measurement.  For example, the H.E.S.S. experiment reported a measurement of the
total spectrum with 30\% systematic errors \cite{hess}.  Since the systematic
errors are mostly due to uncertainties of hadronic interaction models, it is
unclear how much these will be reduced by the time CTA makes a measurement.  For
this simulation, the prior is a "measurement" of the sum of the electron and
positron input spectra with a 30\% error.

Let $a$, $b$, and $c$ be the normalizations of the electron, positron, and
proton components, respectively.  Let $\alpha_i$ be the value in the $i$th pixel
of the flat template corresponding to a uniform intensity for each particle over
the field of view, and let $\beta_i$, $\gamma_i$, and $\delta_i$ be the value in
the $i$th pixel of the templates for the positron, proton, and electron
deficits, respectively.  Then the model for the positive shadow is:
\begin{equation}
  f_i=(a+b+c)\cdot\alpha_i+b\cdot\beta_i+c\cdot\gamma_i
\end{equation}
and the model for the negative shadow is:
\begin{equation}
  g_i=(a+b+c)\cdot\alpha_i+a\cdot\delta_i
\end{equation}
If $y_i$ and $x_i$ are the values of the $i$th pixels in the data (or in this
case simulation) for the two shadows, $\eta$ is a prior measurement of the total
electron and positron spectrum converted into the appropriate units, and
$\sigma(\eta)$ is the error of that measurement, then the quantity to be
minimized is:
\begin{equation}
  \chi^2=\sum\limits_i\left(\frac{(y_i-f_i)^2}{y_i}+\frac{(x_i-g_i)^2}{x_i}\right)+\frac{(a+b-\eta)^2}{\sigma^2(\eta)}
\end{equation}
which can be done with a simple matrix inversion.

\section{Results}

\begin{figure}
  \centering
  \includegraphics[width=0.75\textwidth]{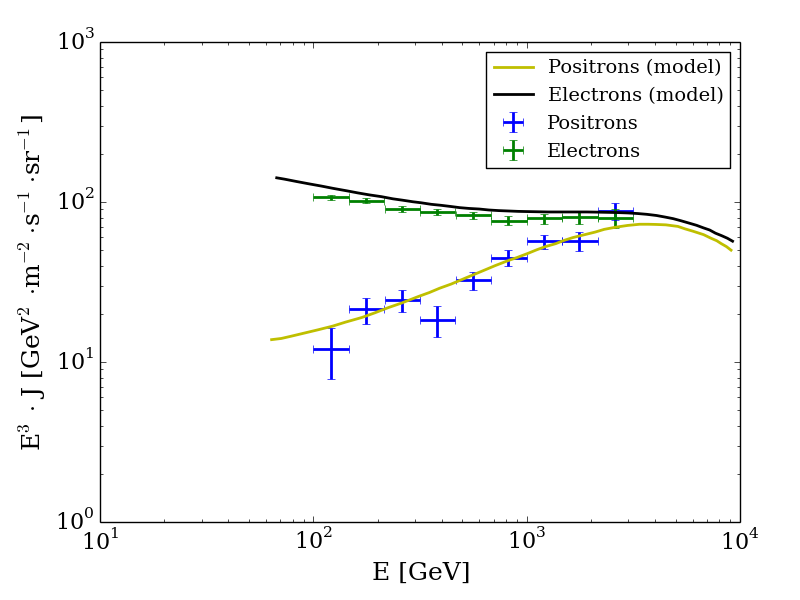}
  \caption{The reconstructed electron and positron spectra using the template
  method and the input spectra.}
  \label{spectra}
\end{figure}

The results of this simulation can be seen in Fig. \ref{spectra}.  From this
plot it is evident that there seems to be a systematic shift in the
reconstructed electron spectrum.  The cause of this shift is under
investigation.  However, the method generally seems to be successful for
reconstructing the input spectra.  The input spectra are "optimistic" in the
sense that the positron fraction is high at TeV energies, but nonetheless the
good performance particularly in the region beyond the current reach of AMS-02
is encouraging.

\section{Future Work}

There are several ways to improve upon the work done so far.  First, the choice
of the field of view for the above work was quite artificial.  A round
8$^{\circ}$ field of view which is fixed on the sky for all energy bins would be
more realistic and would reveal more about the range of energies that could
actually be measured by a CTA observation.  Fortunately, the most interesting
energy bins are the highest energies, for which it will be easier to fit both
shadows in a single field of view.  More study is needed to determine whether it
might be more optimal to divide the observations between two pointings to
capture both shadows over a wider energy range.  As can be seen from the example
templates, it may be unreasonable to extend the measurement down to 100 GeV, as
the shadows could extend to 15$^{\circ}$ from the moon.  However, at that
energy, the positron fraction has already been measured well by AMS-02.

In fact, there is another reason why low energies may not be accessible which is
the high night sky background.  Although the SiPM cameras will not be damaged by
pointing the telescope close to the moon, the increase in the night sky
background rate within the field of view will degrade the reconstruction
performance somewhat and increase the trigger energy threshold.  Detailed
simulations have not yet been done with such high background light conditions.
Future studies with more background light and an updated estimation of the
baseline array layout will help to more accurately characterize the performance,
but again, losing the lowest energies will not be so important.

Finally, the next pass of this study will contain a more accurate treatment of
the proton energy.  It is true that protons and positrons of the same true
energy are deflected by the same amount in the magnetic field, but the events
will be binned in reconstructed energy.  In particular, the proton showers
which are misreconstructed as electromagnetic showers tend to be due to images
of a subshower, and therefore the reconstructed energy is on average about a
third of the true proton energy (with significant spread, again due to the
stochastic nature of the shower development).  This offset has not been taken
into account in this study so far.  Doing this properly could significantly
improve the ability to distinguish protons from positrons because now the
deficits in a reconstructed energy bin will no longer be on top of each other.
In fact, after taking these effects into account, the protons may be better
modeled with a flat distribution rather than with a wide deficit.

With all of the above improvements, this proof-of-concept study will be
developed into an improved estimate of the sensitivity of CTA to the positron
fraction.

We gratefully acknowledge support from the agencies and organizations listed in
this page: http://www.cta-observatory.org/?q=node/22.

\end{document}